\documentclass{xsurv_article}
\usepackage{psfig}

\def\deg{\hbox{$^\circ$}}

\begin{document}

\begin{titlepage}
\setcounter{page}{1}
\makeheadline

\title {Present Epoch Plus - An X-ray survey for Cosmology}

\author{{\sc Keith Jahoda}, Greenbelt, MD USA\\
\medskip
{\small Laboratory for High Energy Astrophysics, Goddard Space Flight Center} \\
}

\date{Received; accepted } 
\maketitle

\summary
This paper summarizes some cosmologically interesting measurements which are uniquely
possible in the hard X-ray band and presents a mission concept capable of achieving them.
The Present Epoch Plus mission will achieve a surface brightness sensitivity of better than
$1\%$ per square degree in the 2-10 keV band,
and create a catalog of $\sim 10^6$ sources.  About 160,000 extragalactic sources are expected
to be detected in the 2-10 keV band, providing an all sky survey with nearly uniform
selection effects 10 times deeper than exisiting or planned surveys.
The PEP concept can be achieved within the size and budgetary constraints of a
NASA Medium Explorer (MIDEX) mission.
END

\keyw
END

\AAAcla
END
\end{titlepage}

%
\kap{Introduction}
Like the Cosmic Microwave Background (CMB), the Cosmic X-ray Background (CXB) above 2 keV
is dominated by an apparently diffuse emission of distant origin.  
The CXB is due to the superposition of numerous unresolved
point sources, and the bulk of the CXB is known to arise from redshifts
$z \ge 1$ even though the details of the source populations remain unknown.  
The CXB provides an ideal observational tool
for studying large scale structure in the universe at redshifts much
greater than those accessible to catalogs of inidvidual objects (which are
always dominated by nearby objects) and at redshifts much much less than the CMB.
Using this tool requires an experiment capable of resolving
and removing the foreground point sources of X-ray emission.
This paper
presents some problems that are uniquely addressed by a mission 
capable of producing 
both precise surface brightness measurements and deep X-ray catalogs,
and sketches a mission concept that achieves these goals.

\kap{Problems Uniquely Addressed by X-ray Missions}
 
Several groups have created plausible models for the origin of the CXB integrating over
the X-ray luminosity functions of various sources, accounting for redshift and evolution
(e.g. Comastri et al. 1995).  The common feature of all models is that the bulk
of the CXB arises between redshifts of 1 and 3.  Using the CXB to study large scale structure
does not depend on the details of the source populations, but only the assumption
that X-ray light traces the underlying matter distribution.  We
highlight a few problems which CXB observations may be uniquely suited to address.

\sect{Velocity with respect to the Baryonic Universe}

The largest anisotropy in the CMB, the dipole with amplitude $\sim 0.001$ of the
mean CMB intensity (Smoot et al. 1992), is widely interpretted as the signature
of peculiar motion, with respect to the CMB, due to gravitiational
acceleration from local anisotropies in the mass distribution.
This interpretation is not controversial;  however, it is central to much of 
cosmology, has not been verified by an independent measurement, is not the only
explanation (Langlois and Piran 1996), and is apparently in conflict with a
measurement of our velocity with respect to a reference frame defined by 
Abell clusters of galaxies (Lauer and Postman 1994).  Confirmation of the peculiar
motion interpretation would come from the measurement of a similar dipole with respect
to distant (i.e. beyond the region producing gravitational acceleration) sources.
Candidate distant sources are the CXB surface brightness distribution and
a population of $> 10^4$ X-ray sources all more
distant than a few 100 Mpc.  Either sample requires separation of
the foreground emission which may be anisotropic due to luminosity associated with the
mass responsible for the peculiar velocity.  The HEAO-1 A2 survey has the statistical
sensitivity to measure the expected dipole (Shafer 1983) but is unable to separate foreground
and background emission (Jahoda 1993). 
Searches for dipole terms in catalogs of point objects
(Maoz 1994, Rauch 1994, Scharf et al. 1995) have fallen an order of magnitude short
of the number of objects required to detect this signal above the shot noise.

\sect{Power Spectrum on scales of 100 - 1000 Mpc and Intermediate Redshift}
Models of structure formation in the universe are challenged to construct models that
simultaneously describe the initial conditions observed by COBE with characteristic
scales of $> 1000$ Mpc and z $\sim$ 1000 and the present day universe observed by
galaxy surveys such as IRAS and which cover scales of a few to a few tens of Mpc
at the present epoch (i.e. z $\sim$ 0) (Fischer et al. 1993).  
Even as the angular scales are pushed closer (MAP and Planck Surveyor will measure the
CMB on sub degree scales while the Sloan Digital Sky Survey (SDSS) will provide
dense samples of galaxies to distances at least ten times greater than IRAS)
the extrapolation
in z remains.  The CXB, if measured on 1 square degree scales, probes z $\sim 1$ and
scales of 100 to 1000 Mpc, helping bridge the gap and providing information at an epoch
which can distinguish several theories of structure formation.  An experiment which also
identifies and removes the foreground sources provides a second chance to measure the power
spectrum at recent epochs.  A measurement made with X-ray selected sources is complementary
to a similar measurement with optically (or otherwise) selected source as X-ray sources
are relatively rare, and thus give better sensitivity at somewhat larger scales.  In
addition, the details of how X-ray sources trace the underlying matter distribution
(often called the bias factor) may be quite different than for other populations,
emphasizing the value of measurements in numerous bands.

\sect{Searches for a Cosmological Constant}
In some critical density cosmologies, some CMB fluctuations arise as late 
$z \sim 1$.  These fluctuations are due to a net Doppler shift experienced by
microwave background photons which traverse expanding potential wells.
This is known
as the Rees-Sciama effect or the integrated Sachs-Wolfe effect, depending on whether the
gravitational perturbation is in the linear or non linear regime (Bennett et al. 1993;
Boughn et al. 1997).
If X-ray light traces the matter distribution and therefore the gravitional potentials, 
one expects a correlation
between X-ray surface brightness and CMB brightness.  Correlations of the HEAO-1 A2 and
COBE DMR maps fail to detect such a signal (Boughn et al. 1997) but the sensitivity is
limited primarily by the size of the beam in both experiments which washes
out small scale fluctuations.  However, the MAP and Planck Surveyor
experiments are being prepared to 
measure CMB fluctuations of scales $< 1 $ square degree.  An X-ray surface brightness
map with similar angular resolution is needed extend this search.

\sect{X-ray catalogs}
Uniformly selected source catalogs in different bands illuminate different aspects of
our universe and discover rather different kinds of sources.  Although ROSAT has discovered
more than $75,000$ sources in its all sky survey (and comparable numbers in the pointed
observations), these source are unlikely to account for the major constituents of the
CXB.  Evidence is increasing that the CXB is dominated by heavily obscured sources which
show up only at energies above the ROSAT band, and which are poorly sampled by the existing
all sky surveys completed with collimated proportional counters.  While many sources
in a 2-10 keV survey will appear in the ROSAT catalog, many others will be new.
ABRIXAS will discover $15,000$ hard band
sources with fluxes above $\sim 10^{-12} \, {\rm erg}\,{\rm s}^{-1}\, {\rm cm}^{-2}$,
about one tenth the number expected from the survey proposed in the following section.
The 2-10 keV catalog will have near uniform sensitivity 
across the whole sky, as the effects of absorption are relatively small.

\kap{The Present Epoch Plus Concept}
We propose a concept capable of performing an all sky surface brightness
survey 
and making the deepest yet X-ray catalog in the
2-10 keV band.  As the primary energy range of interest is $> 2$ keV,
both of these surveys will be relatively unaffected by galactic absorption.
Additionally, the survey will detect more than 5 times as many sources in
the 0.5-2 keV band and make precise surface brightness measurements there
as well.  All of this is accomplished by putting a large number of
ASCA style telescopes into a MIDEX sized payload and scanning the sky for
2-3 years.  
Geometric considerations show that at least 28 ASCA sized telescopes (diameter 38 cm)
can be placed within the 274 cm shroud of a Delta launch vehicle.
Although the detailed engineering has not yet been done, it is clear that
we can use the effective area that would be achieved by 28 ASCA telescopes
is a plausible estimate.  The focal plane instrumentation would not require
28 separate detectors, but could use, for instance, the innovative design
of the ABRIXAS system in which one large 6 by 6 cm CCD is shared among 
seven telescopes.  

We have performed ray tracing simulations using realistic representations
of the ASCA mirrors (K. Gendreau, private communication), representative detector efficiencies for CCDs, and
an exposure which corresponds to a 2 year mission which achieves $70\%$
efficiency (comparable to the achieved ROSAT efficiency;  we might do better
than ROSAT in a lower inclination orbit).  This is equivalent to performing
a series of 817 second exposures with a single ASCA GIS system on a grid
of points separated by $10'$ in each direction.  The input photons were
generated from a randomly distributed collection of sources which obeyed
a Euclidean log N-log S law, each of which had a 1.7 photon index spectrum,
The normalization of the log N-log S power law was 150 sources per square
degree at $10^{-14} \, {\rm erg}\,{\rm s}^{-1}\, {\rm cm}^{-2}$, with the faintest source being determined by the condition
that the total 2-10 keV flux match the observed value ($5 \times 10^{-8} \, {\rm erg}\,{\rm s}^{-1}\, {\rm cm}^{-2} \,{\rm sr}^{-1}$).  The faint sources
supply the bulk of the CXB against which the brighter sources must be detected.
ASCA SIS performance suggests that the CXB will produce 5 times as many photons as the
instrument background.
This simple simulation should be
adequate to assess 2-10 keV sensitivity.
These are
conservative simulations given that the point spread function of foil
mirrors now being produced in our laboratory for ASTRO-E are substantially better
than achieved for ASCA;  improvements will not increase the collecting area, but will
allow smaller detection cells (with lower background rates per cell)
to be used.  
The simulations accepted all photons which start within $2 \deg$ of the 
pointing direction and arrive within $2 \deg$ of the center of the field of
view at the focal plane.  About $30\%$ of all detected photons are assigned
a position in the resulting image $> 30'$ from their celestial point of origin,
so PEP will require a $1 \deg$ collimator in front of the mirrors (or a 
clever baffling system) to eliminate these photons (most of which are single
reflections or reflections off the back of some of the foil segments).  The
simulation produced $\sim 14,000$ photons per square degree without including
the collimators;  we therefore anticipate that the 2 year mission just
meets the $1\%$ goal once the effects of collimators are included.
A sliding box cell algorithm (with $7 \times 7'$ detect cell and $6 \sigma$
significance criteria) detected 48 sources within a 44 square degree region
and is complete to just under
$3 \times 10^{-13} \, {\rm erg}\,{\rm s}^{-1}\, {\rm cm}^{-2}$;  
improvements already demonstrated in mirror point spread function will lower
the flux threshold below $2 \times 10^{-13} \, {\rm erg}\,{\rm s}^{-1}\, {\rm cm}^{-2}$ in accordance with an independant
analytic estimate (R. Warwick, private communication).


\kap{Is a Medium Explorer Necessary?}
A similar experiment concept using only proportianal counters 
and capable of achieving a statistical sensitivity of $1\%$ per 
square degree has been presented by Barcons and collaborators
(Barcons et al. 1997, Barcons 1997).  Although a relatively
large collecting area is required, this experiment can plausibly fit
within the constraints of the NASA SMall EXplorer (SMEX) program.
Barcons et al. (1997) make the case for why $1\%$ per square degree
is a desirable sensitivity.  Briefly, measuring the power spectrum
requires constructing the P(D) curve (the distribution of observed
surface brightnesses) and comparing the measured distribution to
a predicted one.  The prediction can (in the post AXAF era) include
precise information about the log N-log S distribution as well as
details about exposure and statistical sensitivity.  Deviations
from the prediction carry information about effects not included,
such as clustering of sources.  Sensitive measurements of the
clustering (from which the power spectrum can be derived) are limited
by the the total number of independent measurements (which argues for 
a small beam) and the statistical precision of each measurement
(which argues for a large beam if exposure and area are fixed).
The $1\%$ per square degree figure of merit provides a statistical
sensitivity substantially smaller than the expected fluctuations
(of order $10\%$ for 1 square degree beams) with a large number of independent measurements.
Our concept requires a larger collecting area simply because focussing
optics are much less efficient than collimated proportional counters.
Particularly above $\sim 2$ keV, where atomic edges leave a
noticable drop in the efficiency of all astronomical X-ray mirrors,
the net efficiency is $< 25\%$ compared to the proportional counter,
requiring large raw collecting area.   It is necessary to then ask,
what benefits are derived from the larger (and more costly) concept
presented here.  The advantages are:

(a)  This concept produces a contemporaneous X-ray catalog with 
an estimated 150,000 extragalactic sources of flux $> 2 \times 10^{-13} \, {\rm erg}\,{\rm s}^{-1}\, {\rm cm}^{-2}$ in the
2 -10 keV band and more than 600,000 sources independently identified
in the band below 2 keV.  

(b)  The contemporaneous survey allows direct removal of the brighter sources 
from the P(D) curve, increasing sensitivity to the power spectrum at intermediate
redshift.  The proportional counter experiment must mask pixels identified
from a non contemporaneous survey, thus losing signal and introducing the additional
uncertainty of whether the sources at the catalog limit were in fact bright when the
surface brightness measurements were made.  Barcons et al. (1997) quantify this
uncertainty in terms of typical variability of sources with mean flux of
$10^{-12} \, {\rm erg}\,{\rm s}^{-1}\, {\rm cm}^{-2}$ but this variability is in fact not well known.

(c)  The imaging survey allows the brighter sources to be removed without masking
an entire 1 square degree sky pixel.  Not only does this allow the removal of fainter
sources, but the sources can be removed without masking entire pixels which would
substantially reduce sensitivity.

(d)  The catalog allows an independent measurement of the power spectrum in the
present epoch.

(e)  The focussing concept will have much mcuh better sensitivity for searches limited
by shot noise, such as a spherical harmonic decomposition (Lahav et al 1997) or
a cross correlation with the CMB (Boughn et al. 1997).

\kap{Comparison with other missions}
The best all sky surface brightness survey in the 2-10 keV band is still the HEAO-1 A2
survey (Shafer 1983, Boldt 1987, Allen et al. 1994).  The A2 experiment achieved fluctuation
dominated sensitivity, but only on scales of $\ge 5$ square degrees;  the 
only uniform and contemporaneous
source catalog is derived from the same collimated proportional counter experiment
and reaches only $3 \times 10^{-11}\, {\rm erg}\,{\rm s}^{-1}\, {\rm cm}^{-2}$ (Piccinotti
et al. 1982).  Although this survey continues to be of interest for cross correlation
with various other surveys (e.g. the CMB (Boughn et al. 1997) and the soft 
X-ray background (Miyaji et al. 1996)) the limited angular resolution now limits the sensitivity.
The Piccinotti et al. (1982) survey will soon be replaced as the deepest complete
X-ray survey in the band above 2 keV by the ABRIXAS mission (Trumper 1997).  PEP will
detect an order of magnitude more sources than ABRIXAS, but more importantly, has significant
sensitivity to surface brightness.  ABRIXAS will detect 0.11 and 0.34 count/sec over the field of view
from the
CXB and the instrument background, respectively, in the 2-10 keV band (P. Friedrich, private communication).
In a 3 year mission with $70\%$ efficiency, ABRIXAS will achieve a precision of only $15\%$ per square
degree.  The increase in signal to noise for PEP compared
with ABRIXAS is largely due to the increased efficiency achievable with the longer (4 m vs 2 m)
optical bench.  While the possibilities opened by ABRIXAS source catalog will take many years to exploit, that experiment will not contribute to any of the experiments
requiring surface brightness measuremnts.  The combination of deep catalogs and surface brightness
measurements give PEP unique capabilities.

\vspace{0.5cm}

{\it Acknowledgements}: I thank the many colleagues with whom I have discussed these ideas, at the Potsdam
Workshop on X-ray Surveys and elsewhere:  particularly 
X. Barcons, K. Black, E. Boldt, 
S. Boughn, A. Fabian,
K. Gendreau, G. Hasinger, S. Holt, O. Lahav, R. Mushotzky, W. Sanders, 
P. Serlemitsos, M. Treyer, R. Warwick, and N. White.

%
\refer
\aba
\rf{Allen J. et al. 1994, Legacy, 5, 27}
\rf{Barcons X. 1997, this volume}
\rf{Barcons X. et al., 1997, MNRAS, in press}
\rf{Bennett C. et al. 1993, ApJLett, 414, L76}
\rf{Boldt E. 1987, PhysReports, 146, 215}
\rf{Boughn S. et al. 1997, ApJ, submitted}
\rf{Comastri A. et al. 1995, AstronAstrophys, 296, 1}
\rf{Fischer K. et al. 1993, 402, 42}
\rf{Friedrich et al. 1996, MPE Report 263, p. 681}
\rf{Jahoda K. 1993, AdvSpaceRes, 13(12), 231}
\rf{Lahav O. et al. 1997, MNRAS, 284, 499}
\rf{Langlois D. and Piran T. 1996, PhysRevD, 53(6), 2908}
\rf{Lauer T. and Postman M. 1994, ApJ, 425, 418}
\rf{Margon B. 1997, this volume}
\rf{Maoz E. 1994, ApJ, 428, 454}
\rf{Miyaji T. et al. 1996, AA, 312, 1}
\rf{Piccinotti G. et al. 1982, ApJ, 253, 485}
\rf{Rauch M. 1994, MNRAS, 271,13}
\rf{Scharf C. et al. 1995, ApJ, 454, 573}
\rf{Shafer R. 1983, PhD thesis, U.MD.}
\rf{Smoot G. et al. 1992, ApJLett, 396, L1}
\rf{Trumper J. 1997, this volume}
\abe

%
\addresses
\rf{Keith Jahoda, NASA/GSFC Code 662, Greenbelt, MD 20771, USA, e-mail: keith.jahoda@gsfc.nasa.gov}
END
%

\end{document}